\documentclass[aps,prc,twocolumn,superscriptaddress,showpacs]{revtex4}

\usepackage{graphicx}
\usepackage{latexsym}
\usepackage{bm}
\usepackage{amsmath}

\usepackage{color}

\newcommand{\comment}[1]{}

\newcommand{\lsim}{\mbox{\raisebox{-0.6ex}{$\stackrel{<}{\sim}$}}\:}

\newcommand{{\vp}}{{\vec p}}
\newcommand{{\vq}}{{\vec q}}

\newcommand{\beq}{\begin{equation}}
\newcommand{\eeq}[1]{\label{#1} \end{equation}}
\newcommand{\half}{{\textstyle \frac{1}{2} }}
\newcommand{\lton}{\mathrel{\lower.9ex \hbox{$\stackrel{\displaystyle
<}{\sim}$}}}
\newcommand{\gton}{\mathrel{\lower.9ex \hbox{$\stackrel{\displaystyle
>}{\sim}$}}}
\newcommand{\ee}{\end{equation}}
\newcommand{\ben}{\begin{enumerate}}
\newcommand{\een}{\end{enumerate}}
\newcommand{\bit}{\begin{itemize}}
\newcommand{\eit}{\end{itemize}}
\newcommand{\bc}{\begin{center}}
\newcommand{\ec}{\end{center}}
\newcommand{\bea}{\begin{eqnarray}}
\newcommand{\eea}{\end{eqnarray}}
\newcommand{\beqar}{\begin{eqnarray}}
\newcommand{\eeqar}[1]{\label{#1}\end{eqnarray}}

\begin{document}

\title{
Perfect Fluidity of the Quark Gluon Plasma Core\\
as Seen through its Dissipative Hadronic Corona
}

\author{Tetsufumi Hirano}
\affiliation{Department of Physics,
   Columbia University, New York, NY 10027, USA}
\affiliation{RIKEN BNL Research Center,
        Brookhaven National Laboratory,
    Upton, NY 11973, USA}

\author{Miklos Gyulassy}
\affiliation{Department of Physics,
   Columbia University, New York, NY 10027, USA}

\date{\today}
 
\begin{abstract}
  The agreement of hydrodynamic predictions of differential elliptic
  flow and radial flow patterns with Au+Au data at $\sqrt{s_{NN}}=200$
  GeV is one of the main lines of evidence suggesting the nearly
  perfect fluid properties of the strongly coupled Quark Gluon Plasma,
  sQGP, produced at RHIC.  We study the sensitivity of this conclusion
  to different hydrodynamic assumptions on hadro-chemical and thermal
  freezeout after the sQGP hadronizes.  We show that if chemical
  freezeout occurs at the hadronization time, as required to reproduce
  the observed hadron yields, then, surprisingly, the {\em differential}
  elliptic flow,
  $v_2(p_T)$, for pions
   continues to {\em increase} with proper
  time in the late hadronic phase until thermal freezeout and leads to a
  discrepancy with the $v_2(p_T)$ data.  
  In contrast, if both
  hadro-chemical and thermal equilibrium are maintained past the
  hadronization point, then the mean transverse momentum per pion
  increases in a way that accidentally preserves $v_2(p_T)$ from the
  sQGP phase in agreement with the data, but at the cost of
  the agreement with the observed hadronic yields.  In order that
  all the data on (1) hadronic ratios, (2) radial flow, as well as (3)
  differential elliptic flow be reproduced, the sQGP core 
  must expand with a minimal viscosity, $\eta \approx T_c^3$,
  that is however even greater than
the viscosity, $\eta_H \approx T/\sigma_H$, of its hadronic
  corona.  However, because of the large entropy density difference of
  the two phases of QCD matter, the larger viscosity in the sQGP phase
leads to 
 nearly perfect fluid flow in that phase
while the smaller entropy density of the hadronic corona strongly
hinders the applicability of Euler hydrodynamics in that phase.
The ``perfect fluid'' property of the sQGP is thus not due to a sudden
  reduction of the viscosity at the critical temperature $T_c$, but to
  a sudden increase of the entropy density 
of QCD matter and is therefore an
important signature of deconfinement.
\end{abstract}


\maketitle

\section{Introduction}
\label{sec:intro}

One of the most intriguing experimental findings at the
Relativistic Heavy Ion Collider (RHIC) in Brookhaven National
Laboratory (BNL) is the large magnitude
of the elliptic flow parameter $v_2$
\cite{Ackermann:2000tr,Adcox:2002ms,Back:2002gz}
in comparison with the smaller values 
observed at lower collision energies (for
results at Super Proton Synchrotron (SPS) energies, see
Refs.~\cite{Alt:2003ab,Agakichiev:2003gg,Aggarwal:2004ub}).  The
magnitude of $v_2$ and in particular 
its transverse momentum $p_T$ and mass $m$
dependences at RHIC were found 
to be close to predictions based on \textit{ideal},
non-dissipative hydrodynamics simulations around midrapidity ($\mid
\eta \mid \lsim 1$), in the low transverse momentum region ($p_T \lsim 1$
GeV/$c$), and up to semicentral collisions ($b \lsim 5$ fm)
\cite{Kolb:2000fh,Hirano:2001eu}.
This result has led to
the recent BNL announcement \cite{BNL}
 about the discovery of the near perfect fluidity
 of the strongly coupled/interacting quark gluon plasma
(sQGP) \cite{Lee:2005gw,Gyulassy:2004zy,Shuryak:2004cy}
produced in ultra-relativistic nuclear reactions at RHIC. 

Until RHIC data, ``perfect fluidity'' was never observed
nor expected  to apply theoretically in high energy 
hadronic or nuclear reactions
due to nonvanishing viscous dissipation \cite{namiki}.
Especially, since the discovery
of asymptotic freedom in QCD, the prevailing 
paradigm has been the expectation of large viscosities
in a weakly
coupled/interacting QGP (wQGP) at very high densities.
In addition, it is well established \cite{Stoecker:2004qu}
that the hadronic resonance
gas phase of QCD matter is highly dissipative.
The discovery of elliptic flow at RHIC consistent
 with nearly perfect fluidity  
is therefore an experimental and theoretical  surprise. 
Hence a new name, sQGP, has been adopted to characterize
the observed strong coupling properties of the QGP
near the critical temperature $T_c\sim$ 160--170 MeV
that keep viscous effect to a minimum at RHIC.

In this paper, we present the case for the following 
physical interpretation of RHIC data based
on current hydrodynamic
analyses: (1) the high density core part of matter produced 
in relativistic heavy ion collisions, \textit{i.e.} the sQGP,
must expand as a nearly perfect fluid
despite of its \textit{higher} viscosity,
(2) the perfect fluidity of the sQGP core is a consequence
from a large jump of the entropy density at the
critical temperature, $T_c$, \textit{i.e.} deconfinement,
and not from some anomalous reduction of its viscosity,
(3) viscous effects on its hadronic corona
are necessarily large despite its {\em smaller} viscosity,
and (4) ideal inviscid hydrodynamics should not be applied to
the hadronic corona which requires
a nonequilibrium  transport description.

In Sec.~\ref{sec:etaovers}, 
we discuss why we expect a surprising monotonic increase of
the viscosity of QCD matter through the critical temperature
and emphasize the important role played by the rapidly varying viscosity
to entropy {\em ratio} in connection with
perfect fluidity of the sQGP phase.
In Sec.~\ref{sec:hydroissue},
we discuss different assumptions
for the hadronic matter in the hydrodynamic models
to clarify what are open issues
in the current hydrodynamic approaches.
In Sec.~\ref{sec:ET}, the time evolution of the transverse
energy per particle is discussed.
The mean transverse energy is found to be the key
to distinguish the model assumptions in the hadron phase.
Results from the hydrodynamic simulations are
reviewed in Sec.~\ref{sec:hydro}.
We will show how the perfect fluid description
for the hadronic matter in chemical equilibrium
in the conventional hydrodynamic simulations
leads to accidental reproduction of $p_T$ spectrum
and $v_2(p_T)$. 
In order to understand analytically the role of chemical freezeout
on the transverse dynamics,
we employ a blast wave model and give a dynamical meaning
to this model in Sec.~\ref{sec:analytic}.
Finally,
summary of this study and an outlook are presented in
 Sec.~\ref{sec:summary}.

\section{Viscosity and Entropy in QCD}
\label{sec:etaovers}

Weak coupling perturbative QCD (pQCD) 
estimates \cite{Hosoya:1983xm,Danielewicz:1984ww,Thoma:1991em}
 of the viscosity of a wQGP were based on basic kinetic theory
relations
\begin{eqnarray}
\label{wqgp}
 \eta_{\mathrm{wQGP}}
&\approx &
\frac{4}{15} \epsilon_{\mathrm{SB}}(T) \lambda_{\mathrm{tr}}% \nonumber \\
 \approx  \frac{1}{5}\frac{T}{\sigma_{\mathrm{tr}}} 
\frac{s_{\mathrm{SB}}(T) }{n_{\mathrm{SB}}(T)},
 \nonumber \\
\frac{\eta_{\mathrm{wQGP}}}{s_{\mathrm{SB}}}  & \approx & 
\frac{T\lambda_{\mathrm{tr}}}{5}
\end{eqnarray}
where (in $\hbar=c=k_B=1$ units), 
 $\epsilon_{\mathrm{SB}}(T)=3P_{\mathrm{SB}}(T)=\frac{3}{4}T s_{\mathrm{SB}}(T)
\approx 3 T n_{\mathrm{SB}}(T) \approx K_{\mathrm{SB}} T^4 $ 
is the energy density, pressure, entropy density, and number density 
of an
ideal Stefan-Boltzmann (SB) gas of quarks and gluon
characterized by the constant 
$K_{\mathrm{SB}}=\frac{\pi^2}{30}[2(N_c^2-1)+ \frac{7}{8} 12 N_f]
\sim 12$--15 for $N_c=3,N_f=2$--3.
The key microscopic dynamical quantity in Eq.~(\ref{wqgp})
 is the 
transport mean free path $\lambda_{\mathrm{tr}}=1/(n_{\mathrm{SB}}\sigma_{\mathrm{tr}})$
which
is controlled in pQCD by the Debye screened transport 
cross section \cite{Danielewicz:1984ww,Molnar:2001ux}
\begin{eqnarray}
\label{sig}
\sigma_{\mathrm{tr}}&=& \int d\sigma_{\mathrm{el}}\; 
\sin^2\theta_{\mathrm{cm}}
\nonumber \\
& =& \frac{8\pi\alpha_s^2}{\hat{s}}(1+z)
\left[(2z+1)\ln\left(1+\frac{1}{z}\right)-2\right]
\; \; , \label{sigtr} \end{eqnarray}
where $z=\mu^2/\hat{s} $  and $\hat{s}\approx 17 T^2$
is the mean partonic Mandelstam variable. Perturbatively, the screening mass
squared varies as $\mu^2\approx 4 \pi \alpha_s T^2$. For numerical estimates
we take $\alpha_s(T)\approx 4\pi/[18 \ln(4 T/T_c)]$, so that 
$\alpha_s(T_c)\sim 0.5$. 
In the range $T_c< T< 5 T_c$ ($0.5 >\alpha_s(T) > 0.23$),
the perturbative transport cross section
$\sigma_{\mathrm{tr}}< 2 $ mb remains 
much smaller than typical
hadronic cross sections 
$\sigma_H\sim 10-20$ mb
\cite{Danielewicz:1984ww,Gavin:1985ph,Muronga:2003tb}.

An important dimensionless measure
of how imperfect or dissipative a 
fluid may be given by
the ratio of viscosity to entropy density, $\eta/s$ \cite{noteoncausal}. 
This is most easily seen via the Navier-Stokes equation
in (1+1)-dimensional boost invariant
hydrodynamics \cite{Hosoya:1983xm,Danielewicz:1984ww,Gavin:1985ph}. 
In the perfect (Euler) fluid limit, 
the proper energy density decreases with proper 
time, $\tau$, 
 due to longitudinal expansion $dV=\pi R^2d\tau$ and $PdV$ 
work 
via $d\epsilon/d\tau
=-(\epsilon+P)/\tau=-s T/\tau$
with a solution $T=T_0(\tau_0/\tau)^{1/3}$
for massless particles \cite{Bjorken:1982qr}. 
However,
 shear and bulk viscosity reduce the ability of the system 
to perform useful work by adding a term $(4\eta/3+\zeta)/\tau^2$ 
to the right hand side. Neglecting bulk viscosity, $\zeta$,
that vanishes in equilibrium for massless partons,
$d\epsilon/d\tau=-(sT/\tau)[1-4(\eta/s)/3T\tau]$. This
 shows that for the earliest 
times consistent with the uncertainty principle \cite{Danielewicz:1984ww},
$\tau\sim 1/3T$, the cooling of the plasma
 due to both expansion and work
is canceled if $\eta/s > 1/4$. The ability of the system to convert
internal energy into collective flow is thus severely 
impaired at early times if $\eta/s$ and $\zeta/s$ are
 not very small. In fact, in 
order to reproduce the observed 
elliptic flow at RHIC, numerical solutions
to covariant parton transport equation \cite{Molnar:2001ux}
and blast wave analysis with viscous corrections
 \cite{Teaney:2003pb}
showed that
 $\eta/s$ had to be
less than about 0.2 during the first
3 fm/$c$.

The viscosity to entropy ratio in the weakly coupled QCD plasma  on the other
hand is 
\begin{equation}
\label{weakqgp}
\left( \frac{\eta}{s}\right)_{\mathrm{wQGP}}
= \frac{3}{5}\frac{T}{\sigma_{\mathrm{tr}}}
\frac{1}{K_{\mathrm{SB}} T^3} 
\approx \frac{0.071}{\alpha_s(T)^2 \ln[1/\alpha_s(T)]}
\end{equation}
This ratio is not small
($\eta/s=0.35, 0.48, 0.58,0.66$ for  $T/T_c =1,1.5,2.0,2.5$)
indicating that the wQGP is expected to be a rather 
``poor fluid'' with large dissipative corrections. 
 
The analytic dependence on $\alpha_s$ in Eq.~(\ref{weakqgp})
reproduces well 
the approximate temperature dependence implied by Eq.~(\ref{sig})
if we assume the perturbative variation of the screening scale $\mu$.
Lattice QCD data \cite{Kaczmarek:1999mm} indicate 
that $\mu\approx$ (2.0--2.5)$T$ 
is not far from the perturbative estimate extrapolated into the 
physical $g>1$
region and that $\alpha_s(T) < 0.5$ above $T_c$ is also reasonable.
However, the underlying simple gas kinetic approximation for the viscosity
is only rigorously valid in the $g \ll 1$ region. 

Formally, by increasing $\alpha_s>0.5$, it would seem that 
 the right hand side of Eq.~(\ref{weakqgp})  
 could be made to be as small as we like if we ignore the
$\ln(1/\alpha_s)$ singularity.
However, by the Heisenberg uncertainty principle,
the transport mean free path cannot be localized to
less than $\Delta x \sim 1/\langle p\rangle \sim 1/3 T$.
 This leads to 
a quantum kinetic lower bound on the viscosity
for ultrarelativistic gases \cite{Danielewicz:1984ww}:
\begin{equation}
\label{uncert}
\frac{\eta}{s_{\mathrm{SB}}} \gton \frac{1}{15}\;
\end{equation}
with an undetermined multiplicative factor on the order of unity. 

A special quantum field theoretic determination of a 
viscosity lower bound was
found recently
 for \textit{infinitely} coupled supersymmetric Yang-Mills (SYM)
gauge theory using the Anti de-Sitter Space/Conformal Field Theory
(AdS/CFT) duality conjecture \cite{Kovtun:2004de}:
\begin{equation}
\label{ads}
\left( \frac{\eta}{s}\right)_{\mathrm{SYM}} 
= \frac{1}{4\pi} 
\; \; . \end{equation}
This bound is obtained in the  dual $N_c=\infty$ and $g^2N_c=\infty$ limits
of the special ${\cal N}=4$ conformal SYM \cite{schalm}.
It is interesting to note that
this analytic SYM bound is close to the simple 
kinetic theory uncertainty principle bound in 
Eq.~(\ref{uncert}). 
It has been
conjectured \cite{Kovtun:2004de} that $1/4\pi$ in Eq.~(\ref{ads}) 
is the universal minimal viscosity to entropy ratio
 even for QCD. In that case, the viscosity of the sQGP could
be up to a factor of $\sim 1/2\pi$ smaller than of a wQGP. It is then tempting
to conclude that the sQGP must have anomalously small 
viscosity if perfect fluid
behavior is observed. However, as we show below, the sQGP viscosity
is actually very close to that of ordinary hadronic matter just below $T_c$.

To develop this argument further, 
we first digress to recall that the entropy density in the 
$N_c\gg 1, g^2N_c\gg 1$ limits of ${\cal N}=4$ SYM
is given by \cite{Gubser:1998nz}
\begin{equation}
\label{gubser}
s_{\mathrm{SYM}}= \left[ \frac{3}{4}+
\frac{0.6}{(g^2N_c)^{3/2}}+ O\left(\frac{1}{N_c^2}\right)
\right]
\frac{4}{3}K_{\mathrm{SYM}} T^3
\;\; .
\end{equation}
where the Stefan-Boltzmann constant for ${\cal N}=4$ SYM is
$K_{\mathrm{SYM}}={\pi^2 (N_c^2-1)}/{2}\approx 39.5$
is about 3 times greater than 
$K_{\mathrm{SB}}$ of our QCD world \cite{schalm}.
What is especially remarkable about Eq.~(\ref{gubser}) is that,
at infinitely strong  coupling,
 the entropy density is
only reduced by $\sim 25\%$ from its non-interacting SB value.
On the other hand, the viscosity in this extreme limit 
is reduced about an order of magnitude from the weak coupling value
and limited only by the quantum (Heisenberg uncertainty)
 bound on the effective scattering rate.
Current
lattice data on the QCD viscosity near $T_c$ \cite{Nakamura:2004sy}
are with large numerical error
bars between these weak and super strong coupling limits
but the relatively small deviation of the lattice 
entropy density from the SB limit is 
consistent with Eq.~(\ref{gubser}).

%%%%%%%%%%%%%%%%%%%%%%%%%%%%%%%%%%%%%%%%%%%%%%%%%%%%%%%%%%%%
\begin{figure}[t]
\begin{center}
\includegraphics[width=3.3in]{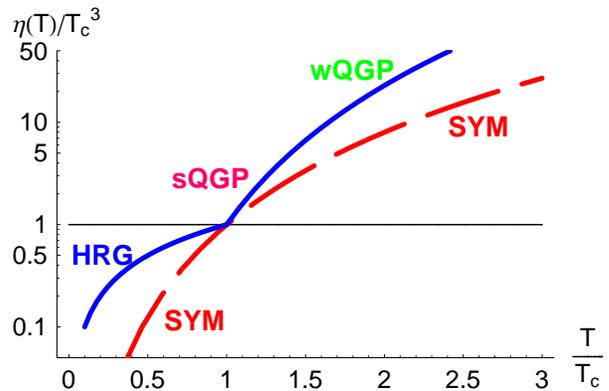}
\caption{Illustration of the approximately monotonic increase of
absolute value of the shear viscosity with temperature.
The kink shown at $T_c$ 
is expected to be smeared out by the $\Delta T_c/T_c\sim 0.1$ 
width of the QCD cross-over transition.
The solid blue curve shows $\eta(T<T_c)=T/\sigma_H$ 
for a HRG followed by the more rapid increase
of the viscosity in the  sQGP phase with  
$\eta_{\mathrm{sQGP}}\approx
 \eta_{\mathrm{SYM}} \equiv K_{\mathrm{SB}}T^3/4\pi \approx T^3$. 
The horizontal line shows that near $T_c$, $\eta\approx\eta_c\equiv T_c^3$.
At high $T\gg T_c$ asymptotic freedom leads
to an even  more rapid growth of viscosity as the sQGP evolves
gradually into the weakly coupled wQGP. In this figure,
 $w=1$ in Eq.~(\protect{\ref{final}})
is taken to emphasize the possibility that the highly viscous
but nearly``perfect fluid'' sQGP
may become an ordinary ``viscous fluid'' already for $T\gton 2 T_c$.
}
\label{fig1a}
\end{center}
\end{figure}
%%%%%%%%%%%%%%%%%%%%%%%%%%%%%%%%%%%%%%%%%%%%%%%%%%%%%%%%%%%%

%%%%%%%%%%%%%%%%%%%%%%%%%%%%%%%%%%%%%%%%%%%%%%%%%%%%%%%%%%%%
\begin{figure}[t]
\begin{center}
\includegraphics[width=3.3in]{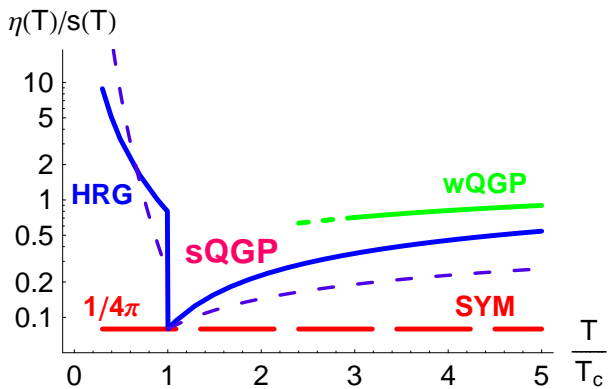}
\caption{Illustration of the rapid variation of the 
\textit{dimensionless ratio}
of the shear viscosity, $\eta(T)$, to the entropy density, $s(T)$.
 The sharp discontinuity illustrated is not due to a rapid change
of the transport coefficient (see Fig.~\ref{fig1a}) but to the 
rapid increase of the entropy density in QCD near $T_c$.
As in Fig.~\ref{fig1a}, we expect the discontinuity to 
be smeared into  a rapid drop within $\Delta T_c/T_c\sim 0.1$. 
Solid (dashed) blue curve
illustrates the change of $\eta/s$ of a HRG with  
$c_H^2=1/3$ (1/6), $s_Q/s_H=10$ (3) into an approximate
 ``perfect fluid'' sQGP at $T_c$. 
The red long dashed curve is $(\eta/s)_{\mathrm{SYM}}=1/4\pi$.
At $T\gg T_c$ asymptotic freedom gradually  transforms
the sQGP into an ordinary viscous fluid wQGP (green), here shown for
 $w=\half, 1$.
}
\label{figeta}
\end{center}
\end{figure}
%%%%%%%%%%%%%%%%%%%%%%%%%%%%%%%%%%%%%%%%%%%%%%%%%%%%%%%%%%%%

The AdS/CFT lower bound (\ref{ads}) together with the assumed 
universal $3/4$ reduction
of the SB entropy density implies that the absolute value of the sQGP viscosity
at $T_c$ would be
\begin{equation}
\label{etasqgp}
\eta_{\mathrm{sQGP}}(T) \approx \eta_{\mathrm{SYM}}(T)
=  
\frac{ K_{\mathrm{SB}} T^3}{4\pi} \approx  T_c^3 \left(\frac{T}{T_c}\right)^3
\end{equation}
where we used a fact that for QCD 
$K_{\mathrm{SB}}\approx
12$--15 is  accidentally close to $4\pi$.
The monotonic increase of $\eta_{\mathrm{SYM}}(T)$
is illustrated by the dashed curve in Fig.1.

The effective transport cross section via Eq.~(\ref{wqgp}) 
at $T_c \sim 160$ MeV is in this case
\begin{equation}
\label{sigc}
\sigma_{\mathrm{tr}}^c
\approx\frac{4}{5}\frac{T_c}{\eta_c} \sim 12 \;{\rm mb}
\; \; . \end{equation}
Here $\eta_c\equiv T_c^3= 0.106 $ GeV/fm$^2$ at $T_c=0.16$ GeV.
See Ref.~\cite{Peshier:2005pp} for an independent
 estimate of the transport cross section in the sQGP phase
leading to similar $\sigma_{\mathrm{tr}}(T)$ near $T_c$.

While there is no consensus yet on what physical mechanisms
could enforce the minimal viscosity bound in the 
sQGP \cite{Molnar:2001ux,Shuryak:2003xe,Xu:2004mz}, we take
as empirical fact that the sQGP
viscosity must be close (within a factor of two)
to the minimal (uncertainty) bound, Eq.~(\ref{etasqgp}).
Our central assumption is that local thermal equilibrium 
is maintained in the sQGP core with minimal dissipative deviations
and with the equation of state and hence speed of sound
as predicted by QCD.
Alternate scenarios, with arbitrary equations of state with
higher  speed of sound that in principle 
could compensate the higher dissipation and viscosity 
in a wQGP will not be considered here. In this connection
we also emphasize the importance of fixing sQGP 
initial conditions with Color Glass Condensate 
or saturating gluon distributions
constrained by the 
global entropy observables
\cite{Gyulassy:2004zy,Hirano:2004rs}.
With fixed initial conditions and equation of state,
the remaining degrees of dynamical freedom are reduced to
the dissipation corrections discussed in this section for the sQGP phase
and the dynamical constraints on its dissipative hadronic corona
discussed in the subsequent sections.

Note that 
the effective transport cross section
in the sQGP $\sigma_{\mathrm{tr}}^c$ just above $T_c$ 
is remarkably close
to the hadron resonance gas transport cross 
section just below $T_c$ \cite{Gavin:1985ph,Muronga:2003tb}.
 However, due to the $1/T^2$ scaling
at $T\sim 2 T_c$, 
the effective transport cross section in the sQGP 
would already drop to $\sim 3$ mb while preserving 
the (uncertainty principle) lower bound Eq.~(\ref{ads}). 

In contrast to the novel sQGP phase above $T_c$, 
for $T<T_c$, matter is well known to be
in the confined hadron resonance gas (HRG) phase
where the kinetic theory viscosity \cite{Danielewicz:1984ww, Gavin:1985ph} is
\begin{equation}
\label{etaH}
\eta_{\mathrm{HRG}}\approx \frac{T}{\sigma_H} \approx \eta_c \frac{T}{T_c}
\; \; , \end{equation}
as illustrated by the solid curve below $T_c$ in Fig. 1.
Because the hadronic transport cross sections
are typically $\sigma_H \sim 10-20$ mb,
the combination of Eqs.~(\ref{etasqgp}) and (\ref{etaH}) shows
that
we should not expect a large variation 
of the absolute value
of the matter viscosity across $T_c$
if the minimal $\eta/s$ holds 
above $T_c$. 
In Ref.~\cite{Gavin:1985ph}, Gavin found that for a pion gas with 
P-wave $\rho$ and D-wave $f^0$ resonance interactions,
the thermal averaged transport cross section 
and viscosity from his Fig.~3 
are $(\eta/\eta_c,\sigma_H [\mathrm{mb}])\sim
(0.66, 20),~(0.9, 17),~(1.1, 15)$  for $T=180,~160,~140$ MeV.
 In Ref.~\cite{Muronga:2003tb}, Muronga used the UrQMD resonance gas
Monte Carlo to compute $\eta(T)/\eta_c \sim 0.75,~1.1,~1.9$
for $T=0.14,~0.16,~0.18$ GeV.
{In both studies \cite{Gavin:1985ph,Muronga:2003tb} }
numerical estimates are thus consistent with  Eq.~(\ref{etaH})
 for $T<T_c$. For nonvanishing baryon density, see recent estimates
 in Ref.~\cite{Muroya:2004pu}.

In the sQGP phase, the minimal viscosity, Eq.~(\ref{etasqgp}),
is predicted 
to grow cubically with $T$ beyond $T_c$. 
However, at $T\gg T_c$ asymptotic freedom predicts
that it would grow even more rapidly as the sQGP transforms
gradually into a wQGP. An interpolation formula
between these phases can be constructed as
\begin{equation}
\label{final}
\eta(T)\approx T_c^3 \left\{ \begin{array}{ll}
(T/T_c)^1 ,& T< T_c \\
(T/T_c)^3[1+w(T)\ln(T/T_c)]^2 ,& T>T_c
\end{array}
\right.
\end{equation}
The extra squared logarithmic growth of the viscosity at $T\gg T_c$
is expected from Eq.~(\ref{weakqgp}). 
To {be} consistent
with the perturbative wQGP at $T\gg T_c$ one should take
\begin{equation}
\frac{w^2(T\gg T_c)}{4\pi} = \frac{9 \beta_0^2}{80\pi^2 K_{\mathrm{SB}}}
\frac{1}{\ln 1/\alpha_s(T)}
\; \; .
\label{c}
\end{equation}
With $K_{\mathrm{SB}}=12$--15 and $\beta_0=11-2N_f/3 \sim 9$--10,
a possible scenario may be that $w\sim 1$ already near $T_c$.
This possibility is shown in Fig.~\ref{fig1a} by the solid curve
above $T_c$ which would imply
sQGP$\rightarrow$ wQGP already above $\sim (2-3)T_c$.
In fact, current lattice 
data on the evolution of screening scales
in the QGP phase suggest that hadronic scale correlations
may persist only up to $T\sim 3 T_c$ \cite{Asakawa:2003re,Datta:2003ww}.
A  value  $w(T>2 T_c) \sim 1$, is also not inconsistent with current
lattice results \cite{Nakamura:2004sy}.
We note that future measurements of elliptic flow 
in A+A collisions
at LHC with $\sqrt{s_{NN}}=5500$ GeV 
will be able to
 test experimentally if such a precocious onset of dissipative wQGP 
dynamics occurs.

The approximate continuity of the viscosity across $T_c$ indicated
in Fig.~\ref{fig1a}
is to be understood to hold within a factor on the order of unity.
What changes rapidly at $T_c$ is not the viscosity
of QCD matter but rather its entropy density due to the deconfinement
of the quark and gluon degrees of freedom.

For a hadronic resonance gas charactered by a speed of sound $c_H^2$,
the entropy density $s(T)=s_H(T/T_c)^{1/c_H^2}$
with decreasing temperature 
decreases much more rapidly than does the viscosity
for typical $c_H^2 \sim 1/6$--1/3.
Just beyond $T_c$  --possibly only up to several times $T_c$-- 
it is postulated 
that the sQGP phase may exist with $\eta/s < 0.2$
but with viscosity close to ideal gas.

Summarizing the discussion up to now, 
we expect that $\eta$ varies smoothly near $T_c$ as in Fig.~\ref{fig1a}
while the ratio $\eta/s$ may have a significant discontinuity
due to the rapid onset of deconfinement as a function of $T$
as shown in Fig.~\ref{figeta}.
We therefore propose the following interpolation
formula for the temperature dependence of the $\eta/s$ ratio 
with a $T$ independent constant $w$ 
\begin{equation}
\label{drop}
\frac{\eta(T)}{s(T)} \approx \frac{1}{4 \pi}
\left\{\begin{array}{ll}
\left( \frac{s_Q}{s_H}\right)
\left(\frac{T}{T_c}\right)^{1-1/c_H^2}, &  T< T_c \\
\left[1+ w\; \ln(T/T_c)\right]^2, & T>T_c \end{array} \right.
\end{equation}
with the negative discontinuity 
\begin{equation}
\left[\frac{\eta}{s}\right]_{T_c}
=\left.\frac{\eta(T_c)}{s(T_c)}\right|_Q
-\left.\frac{\eta(T_c)}{s(T_c)}\right|_H
=-\frac{1}{4\pi}\left(\frac{s_Q}{s_H} -1\right)
\; \; . \label{jump}
\end{equation}
We illustrate Eq.~(\ref{drop})  
in Fig.~\ref{figeta}.
Note that it is the entropy jump $s_Q/s_H\sim 3$--10 
that causes a drop of
$\eta/s$ across $T_c$.  Since the HRG$\rightarrow$ sQGP transition 
with dynamical quarks appears from the lattice QCD 
to be only a
rapid crossover, the discontinuity is understood to be spread out 
over a temperature range $\Delta T_c/T_c \sim 0.1$.

\section{Imperfect Fluidity of the Hadronic Corona}
\label{sec:hydroissue}

In the last section we presented the case that
the $\eta/s$ ratio may be small enough above $T_c$ 
in the sQGP core for perfect fluidity to arise during the critical
first $\sim 3$ fm/$c$,
 while the spatial azimuthal 
asymmetry of the matter produced in non-central reactions is large enough
to induce collective elliptic flow. However,
during the subsequent $\sim 10$ fm/$c$ evolution
after hadronization of the sQGP core, the whole
system evolves as HRG corona. In the HRG, $\eta/s$ 
is too high for local equilibrium to be maintained due to its small
entropy density compared with sQGP.
Nevertheless, 
the data on $v_2(p_T)$ seem to agree very well with some hydrodynamic
predictions based on the assumption that local equilibrium is maintained until thermal freeze-out. However, various assumptions about the 
hadro-chemical evolution are known to lead to significantly different
predictions for the differential elliptic flow.
In this section we focus on 
the question of the validity of the application of hydrodynamics
to analyze the entire evolution in A+A at RHIC. 

The key problem that we now address 
is the role of final state {\em hadronic} interactions in possibly 
modifying conclusions inferred about the prefect fluidity of the sQGP core. 
In order that the sQGP elliptic flow signature of perfect fluidity
survives during the evolution through the extended hadronic ``corona''
we must study how longitudinal flow, transverse radial flow, as well
as the elliptic deformation of the transverse flow may evolve
in hadronic matter.

Several puzzling features suggest the importance of
investigating more closely the distortions that can be caused
by final state hadronic interactions involving 
hadro-chemical and thermal freezeout.
In ideal hydrodynamics it is well known that 
while the central rapidity region is well reproduced by hydrodynamics,
this is not the case 
in forward/backward rapidity regions \cite{Hirano:2001eu}.
Hydrodynamics also strongly overestimates $v_2$
at energies below $\sqrt{s_{NN}}=$200 GeV as well as in the most peripheral
collisions where initially a larger fraction of the transverse
elliptic interaction region starts out in the hadronic phase.
All these data point to the fact that the dynamics in the hadronic
corona is not at all ideal.
 
Another important issue in ideal hydrodynamics as well as in other
dynamical models is the so-called HBT puzzle \cite{Heinz:2002un}.  In
spite of the apparent success of hydrodynamic description for elliptic
flow, hydrodynamics fails to reproduce the experimental data of the
HBT radii \cite{Adler:2001zd,Adcox:2002uc,Back:2004ug}.  The best
current description of hadron freezeout consistent with the HBT puzzle
involves an assumption of a highly dissipative resonance gas dynamics
and transport \cite{Lin:2002gc}.

As compiled in Fig.~20 in Ref.~\cite{Adcox:2004mh},
some hydrodynamic results
reproduce neither $v_2(p_T)$ nor
$p_T$ spectrum.
This immediately raises
the following two questions:

(Q1) Are hydrodynamic results
 consistent with each other at RHIC energies?
What assumptions lead to the differences among hydrodynamic predictions?

(Q2) How robust is the statement that
hydrodynamic description at RHIC data
is good at low $p_T$?

\noindent
The differences 
arise from the treatment of hadron phase
dynamics. The treatment of the sQGP phase is essentially
the same in all approaches so far:
Parton chemical equilibrium and inviscid hydrodynamics
are assumed in the sQGP phase.
There are, however, three generic classes 
of approaches to the 
evolution of the hadronic corona in the literature.

\textit{Chemical equilibrium model (CE).}
Most of the hydrodynamic calculations so far are based on the
assumption that the hadron phase is a perfect fluid
in both hadro-chemical and thermal equilibrium.
With this assumption,
the centrality, transverse momentum, and/or (pseudo)rapidity
dependences of elliptic flow are studied 
\cite{Kolb:2000fh,Hirano:2001eu}.
However, it is known that the yields of heavy hadrons
(essentially all hadrons except for pions)
are smaller in CE than data since the numbers of particle
are counted on
the hypersurface at \textit{thermal} freezeout
within this approach.
At relativistic collisional energies,
thermal freezeout temperature $T^{\mathrm{th}}$
is smaller than chemical freezeout temperature
$T^{\mathrm{ch}}$ \cite{Heinz:1997za,Shuryak:1997xb}.
So the numbers of 
heavy particles are suppressed due to the Boltzmann factor
and, eventually, lead to discrepancy from the data.
CE therefore fails to account for the observed particle abundance
systematics \cite{Braun-Munzinger:2003zd}.

\textit{Partial chemical equilibrium model (PCE).}
In Refs.~\cite{Arbex:2001vx,Hirano:2002ds,Teaney:2002aj,Kolb:2002ve},
chemical freezeout being separated from thermal freezeout
is taken into account in the hydrodynamic simulations
toward simultaneous 
reproduction of particle ratios and particle spectra.
Below $T^{\mathrm{ch}}$,
one introduces chemical potential for each
hadron associated with the conserved number.
Inelastic processes only through
strong interactions are supposed to be equilibrated
in the hadron phase, \textit{e.g.} $\mu_\rho = 2\mu_\pi$,
$\mu_\Delta = \mu_N + \mu_\pi$, and so on.
Note that the conserved pion number within this approach
is $\tilde{N}_\pi = N_\pi + \sum_R b_R N_R$.
Here $b_R$ is the effective branching ratio and $N_i$ is the
number of $i$-th hadron \cite{Bebie:1991ij}.
For details,
see Refs.~\cite{Hirano:2002ds,Teaney:2002aj,Kolb:2002ve}.
This particular model does not reproduce $v_2(p_T;m)$
nor $p_T$ spectra at RHIC
as shown in Fig.~20 in Ref.~\cite{Adcox:2004mh}.
Note that a model is called ``chemical freezeout (CFO)"
when the number of hadrons $N_i$ instead of $\tilde{N}_i$
is conserved. This means even inelastic scatterings 
through strong interaction cease to happen.

\textit{Hadronic cascade model (HC).}
One can utilize a hadronic cascade model
just after the phase transition between
the QGP and hadron phases
\cite{Bass:2000ib,Teaney:2000cw}.
This approach dynamically describes
both chemical and thermal freezeouts 
without assuming explicit freezeout temperatures.
Viscous effects are effectively taken into account 
through the non-zero mean free path among the particles
(see, e.g. Eqs.~(\ref{wqgp}) and (\ref{etaH})).
\begin{center}
\begin{table}[t]
\caption{Summary of hydrodynamic results. 
Hydrodynamic results
of $v_2(p_T)$ and $p_T$ spectra
for pions and protons are compiled in Fig.~20
in Ref.~\cite{Adcox:2004mh}.
CE, PCE, and HC stand for, respectively,
chemical equilibrium, partial chemical equilibrium,
and hadronic cascade.
Thermal freezeout temperature is assumed to be
chosen to reproduce particle spectra
in the model CE.
Note that 
the model CE can reproduce
the shape of $dN/p_Tdp_T$ for protons, not its yield.}
\label{tbl:hydro}
\begin{tabular}{lccc}
\hline
\hline
Observables & model CE & model PCE & model HC \\
\hline
$v_2(p_T;m)$ & yes & no & yes \\
$dN/p_Tdp_T$ & yes & no & yes \\
ratios & no & yes & yes \\
\hline
\hline
\end{tabular}
\end{table}
\end{center}

Hydrodynamic results from the above three classes of hadro-dynamical 
models
are summarized in Table~\ref{tbl:hydro}.
For reviews of hydrodynamic results at the RHIC energies,
see also \cite{Huovinen:2003fa,Kolb:2003dz,Hirano:2004ta}.
As long as the space-time evolution of the hadron
phase is concerned,
the approach HC seems to be the most realistic one
among the three {classes}.
The second best model
should be the model PCE
from the experimental data of particle ratios and spectra.
The models CE and HC reproduce
$v_2(p_T)$ for pions and protons in low $p_T$ regions,
whereas the model PCE
fails. The failure of PCE for $v_2(p_T)$ is particularly
perplexing since the spatial azimuthal asymmetry is mostly gone by the time
hadronization is competed.
In the following sections,
we shall show why the differences between
the models CE and PCE
appears and why the result from the perfect fluid model CE
eventually looks similar to
the one from the highly dissipative model HC.
The key quantities to understand these differences
are found to be
the temporal behavior of the mean transverse momentum/energy
and the ratio of the particle number to the entropy.

\section{Temporal behavior of transverse energy}
\label{sec:ET}

In this section, we briefly review the time evolution
of total transverse energy and transverse energy per particle
in relativistic heavy ion collisions
within a framework of the Bjorken solution for longitudinal
expansion \cite{Bjorken:1982qr}.

Assuming the Bjorken scaling solution,
the time evolution of the entropy density 
is represented by
$s(\tau)  =  s_0 \tau_0/\tau$.
Here $s_0$ is the initial entropy density
and $\tau_0$ is the initial time.
As long as 
a perfect fluid is considered,
the entropy per unit rapidity is conserved 
\begin{eqnarray}
\label{eq:entro}
\frac{dS}{dy} & = & \int d^2x_\perp \tau s(\tau)
               =  A_\perp \tau_0 s_0,
\end{eqnarray}
where $A_\perp$ is the transverse cross section of a nucleus.
Here we assume a smooth function for the equation of state (EOS).
For EOS with $P=c_s^2\epsilon$ $(0<c_s^2<1/3)$,
the time evolution of energy density
becomes
\begin{eqnarray}
\epsilon(\tau) & = &\epsilon_0
\left(\frac{\tau_0}{\tau}\right)^{1+c_s^2}
\end{eqnarray}
where $\epsilon_0$ is the energy density at $\tau_0$.
Thus the transverse energy per unit rapidity 
\begin{eqnarray}
\label{eq:et}
\frac{dE_T}{dy} & = & \int d^2x_\perp \tau \epsilon(\tau)
 =  A_\perp \epsilon_0 \tau_0
\left(\frac{\tau_0}{\tau}\right)^{c_s^2}
\end{eqnarray}
decreases with proper time due to $PdV$ work in spite of
the conservation of entropy \cite{Gyulassy:1983ub,Ruuskanen:1984wv}.
In the following discussion in this section,
we consider three EOS models for pions and
see time evolution of total transverse energy
and transverse energy per pion.

\textit{Massless Pions}.
The number density $n$ is proportional to the entropy density $s$
for massless pions: $n = (d/\pi^2)\zeta(3)T^3$,
$s = d(4\pi^2/90)T^3$.
Inserting $d=3$ and $\zeta(3) \sim 1.2$,
one obtains $s \approx 3.6 n$.
From Eq.~(\ref{eq:entro}),
the number of pions per unit rapidity $dN/dy$ is also conserved.
Therefore the mean transverse mass,
which is identical to the mean transverse momentum
in the massless pion case,
$\langle m_T \rangle = (dE_T/dy)/(dN/dy) \propto (dE_T/dy)$
 decreases with $\tau$ from Eq.~(\ref{eq:et}).

\begin{figure}[t]
\begin{center}
\includegraphics[width=3.3in]{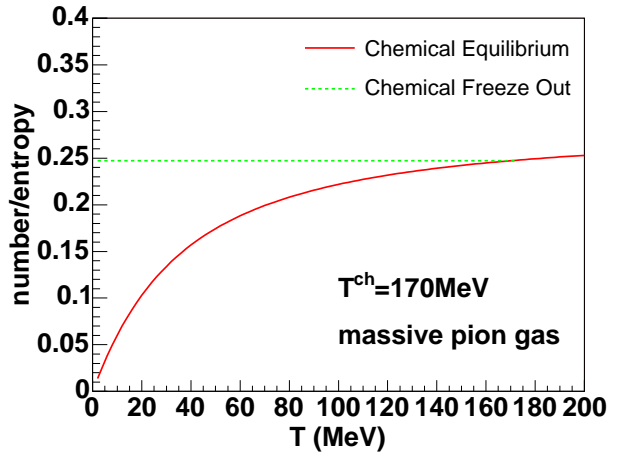}
\caption{
The ratio of number density and entropy density
for massive pions. Solid (Dashed) line
represents the ratio for
chemical equilibrium (chemical freezeout).
Chemical freezeout temperature
is assumed to be $T^{\mathrm{ch}}=170$ MeV.
}
\label{fig:NoverS}
\end{center}
\end{figure}

\textit{Massive Pions in Chemical Equilibrium}.
The proportionality between the number density and the entropy
density is approximately 
valid only for ultra-relativistic particles
($T \gg m$).
In relativistic heavy ion collisions, the typical
temperature is around the order of the pion mass.
So pions are no longer relativistic particles
in this particular situation.
The number density and the entropy density 
for pions are evaluated in the usual prescription
of statistical physics.
Thus the ratio of the number $N$
and the entropy $S$
increases with temperature
due to the finite mass of pions
as shown in Fig.~\ref{fig:NoverS}.
Note that the volume $V$ is canceled and that $N/S$
equals to the ratio of their densities $n/s$.
Therefore $\langle m_T \rangle$ can
\textit{increase} with proper time
(or with decreasing temperature of the system)
even as $dE_T/dy = \langle m_T \rangle dN/dy$ decreases
from Eq.~(\ref{eq:et}).
This ``local reheating'' can occur because
 the total transverse energy is distributed among a smaller
number of pions at lower temperature.
The resulting temporal behavior of $\langle m_T \rangle$
is determined through competition of how rapidly
$dE_T/dy$ and $dN/dy$ decrease with proper time.
As we will see in the next section, $\langle p_T \rangle$
\textit{increases}
with proper time in hydrodynamic simulations with chemical
equilibrium EOS.

%%%%%%%%%%%%%%%%%%%%%%%%%%%%%%%%%%%%%%%%%%%%%%%%%%%%%%%%%%%%
% N/S
%%%%%%%%%%%%%%%%%%%%%%%%%%%%%%%%%%%%%%%%%%%%%%%%%%%%%%%%%%%%
%%%%%%%%%%%%%%%%%%%%%%%%%%%%%%%%%%%%%%%%%%%%%%%%%%%%%%%%%%%%

\textit{Chemically Frozen Massive Pions}.
When the system expands strongly,
inelastic collisions can cease to happen.
So one can think about a situation
in which the system keeps only
thermal equilibrium through
elastic scattering.
Analyses based on statistical models and blast wave models
show that thermal freezeout temperature $T^{\mathrm{th}}$
is smaller than chemical freezeout temperature $T^{\mathrm{ch}}$
at AGS, SPS and RHIC energies.
Moreover, $T^{\mathrm{ch}}$ is found to be close to 
the (pseudo)critical temperature $T_c$.
This indicates that the hadron phase in relativistic heavy
ion collisions is in thermal equilibrium, not in chemical
equilibrium \cite{Heinz:1997za,Shuryak:1997xb}.
Usually, the term ``chemical equilibrium" is associated with
a state equilibrated among a finite number ($>1$)
 of compositions in the system.
Here we simply use the term ``chemical freezeout"
in spite of one hadronic species, \textit{i.e.} pions.
This means the number of pions per unit rapidity
is fixed below $T^{\mathrm{ch}}$.
The entropy is also conserved as long as a perfect fluid
is considered,
so the ratio of the number density and the entropy density
is a constant of motion similar to the case for massless
pions.
It is interesting to mention that the entropy is not generated
even in the evolution of chemically frozen fluids.
Entropy production originates from
the source term in the rate equation
for chemical non-equilibrium processes.
The number of hadron is, however, conserved
after chemical freezeout.
It is easy to show the conservation of entropy
$\partial_\mu S^\mu = 0$
from the conservation of energy and momentum
$\partial_\mu T^{\mu\nu}=0$, 
where $T^{\mu\nu} = (\epsilon +P)u^\mu u^\nu - P g^{\mu\nu}$, and 
the conservation of the number of hadrons 
$\partial_\mu N_i^{\mu}= 0$.
In this sense, one needs to
distinguish ``chemical freezeout" from
``chemical non-equilibrium".
In the chemical non-equilibrium process,
the system is \textit{approaching} to chemical equilibrium
state, \textit{i.e.} the maximum entropy state
 through inelastic scattering.
Entropy is certainly generated during this process.
Contrary to this, chemical freezeout
means that the system suddenly \textit{leaves} from
the chemical equilibrium state
by keeping particle ratios
due to the strong expansion.

Figure \ref{fig:NoverS} shows comparison
of the ratio of pion number $N$
and its entropy $S$ between chemical equilibrium EOS and
chemically frozen EOS.
Here chemical freezeout temperature is taken as
being $T^{\mathrm{ch}}=170$ MeV.
Similar to the massless pion case, $\langle m_T \rangle$
in the chemical freezeout case
 decreases with decreasing decoupling temperature.
As long as the Bjorken scaling solution for
longitudinal expansion is considered, transverse expansion
does not spoil the above discussion: $PdV$ work
done in the transverse direction is simply
converted into the kinetic energy of fluid elements.
The resultant slope of $p_T$ spectrum for pions should become softer
at lower decoupling (thermal freezeout) temperature.
Quantitatively, the $p_T$ slope is insensitive to the choice
of $T^{\mathrm{th}}$ since $dE_T/dy \propto \tau^{-c_s^2}$
changes only gradually in the late stage.
The universal behavior of the $p_T$ slope
is already confirmed in the hydrodynamic
simulations with chemically frozen (or partial chemical 
equilibrium) EOS \cite{Hirano:2002ds,Kolb:2002ve} 
and will be mentioned in the next section.

From the above consideration,
the key quantity which governs the transverse dynamics,
particularly the time evolution of mean transverse mass,
within ideal hydrodynamics 
is found to be the ratio of the number $N$
and the entropy $S$.
It is commonly expected that
the behavior of the mean transverse energy/momentum
is governed by the stiffness of the EOS.
But it is not always true since the sound velocity of
chemical freezeout EOS (or simply the slope of $P$
as a function of $\epsilon$)
is almost the same as that
of chemical equilibrium EOS as shown in Ref.~\cite{Hirano:2002ds}.
Interestingly, whether the \textit{mean transverse energy}
increases or decreases with the proper time
is governed by $N/S$
and the \textit{longitudinal} work,
not the stiffness of EOS.
We will also mention these hydrodynamic results in the next section.

To summarize this section,
$\langle m_T \rangle$ decreases with proper time 
for massless pions or chemically frozen massive pions, while
it can increase for massive pions in chemical equilibrium. 
We emphasize here that these conclusions 
are obtained from quite basic
assumptions:
the first law of thermodynamics ($PdV$ work) and the Bjorken
scaling solution.

\section{Results from Hydrodynamic Simulations }
\label{sec:hydro}

We compare the hydrodynamic results from the model PCE 
with the ones from the model CE
with respect to the EOS, space-time
evolution, $p_T$ spectra, and $v_2(p_T)$.
Hydrodynamic simulations have already been performed
for Au+Au collisions at $\sqrt{s_{NN}}=130$ GeV
and the essential results
have already been obtained in Ref.~\cite{Hirano:2002ds}.
In this section, we make an interpretation of these
hydrodynamic results obtained so far.
In particular, we emphasize that the temporal behavior of the 
mean $p_T$ is the key to understand
the difference of the results 
between these two models.
For further details of the hydrodynamic model,
see also Ref.~\cite{Hirano:2002ds}.

\subsection{Equation of state and space-time evolution}
Chemical freezeout does not change EOS,
\textit{i.e.}~pressure as a function
of energy density $P(\epsilon)$,
so much \cite{Hirano:2002ds,Teaney:2002aj}. This means that
the difference of chemical composition in the hadron phase
does not affect the space-time
evolution of \textit{energy density} significantly.
However, at a fixed temperature, the energy density
in chemically frozen (or partial chemical equilibrium)
hadronic matter
is considerably larger than the one
for hadronic matter in chemical equilibrium.
This is due to the fact that the large resonance
population keeps in the system during the expansion
after chemical freezeout
and that the mass terms 
significantly contributes to the energy density. 
Therefore the space-time evolution of \textit{temperature field}
does change significantly
while $P(\epsilon)$ remains essentially  unchanged.
The temperature of the chemically frozen system cools down
more rapidly than that of the chemical equilibrium system.
This leads to the reduction of life time of fluids,
radial flow, and
HBT radii ($R_{\mathrm{long}}$ and $R_{\mathrm{out}}$) 
\cite{Hirano:2002ds}.
Longitudinal size $R_{\mathrm{long}}$,
which relates with the lifetime of a fluid
through the gradient of
longitudinal flow velocity $(dv_z/dz)^{-1}\approx\tau_f$
\cite{Makhlin:1987gm},
can be interpreted by the effect of
chemical freezeout on the life time of a fluid
in hydrodynamics.
Nevertheless, $R_{\mathrm{side}}$ and $R_{\mathrm{out}}$
are still inconsistent with data.

\subsection{$p_T$ spectra and elliptic flow}

In hydrodynamic simulations with chemical
equilibrium, thermal freezeout temperature
$T^{\mathrm{th}}$ is an adjustable parameter.
In order to fix $T^{\mathrm{th}}$,
one usually utilizes the experimental
data of $p_T$ spectra for hadrons.
Reduction of $T^{\mathrm{th}}$
leads to generation of additional radial flow.
Generically,
the resulting $p_T$ spectra
at $T^{\mathrm{th}}=T_1$
becomes harder
than the ones at $T^{\mathrm{th}}=T_2>T_1$
even though temperature, \textit{i.e.} the inverse slope
of the momentum distribution 
in the local rest frame
decreases.
Contrary to this behavior, 
when chemical freezeout is appropriately
taken into account in hydrodynamic simulations,
the $p_T$ slope
becomes insensitive to $T^{\mathrm{th}}$
compared to the one in the model CE \cite{Hirano:2002ds}.
%%%%%%%%%%%%%%%%%%%%%%%%%%%%%%%%%%%%%%%%%%%%%%%%%%%%%%%%%%%%
% Mean pT
%%%%%%%%%%%%%%%%%%%%%%%%%%%%%%%%%%%%%%%%%%%%%%%%%%%%%%%%%%%%
\begin{figure}[t]
\begin{center}
\includegraphics[width=3.3in]{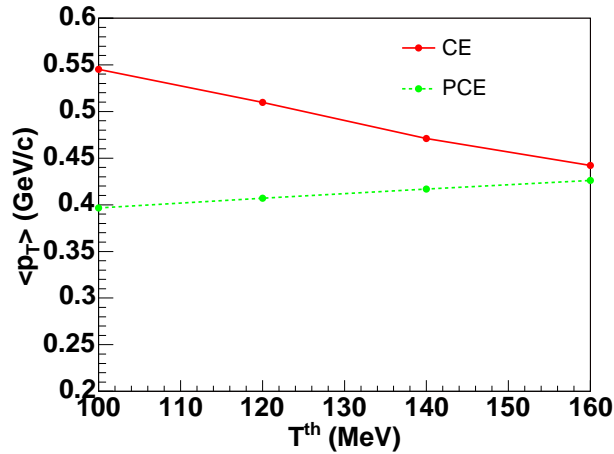}
\caption{
Average transverse momentum for pions as a function of
thermal freezeout temperature 
for the models CE (solid) and PCE (dashed).
}
\label{fig:meanpt}
\end{center}
\end{figure}
%%%%%%%%%%%%%%%%%%%%%%%%%%%%%%%%%%%%%%%%%%%%%%%%%%%%%%%%%%%%

To confirm these behaviors,
we perform hydrodynamic simulations again
for a particular choice of impact parameter $b=5$ fm and
see the average transverse momentum $\langle p_T \rangle$.
Details of the hydrodynamic models used here can be found
in Ref.~\cite{Hirano:2002ds,Hirano:2003pw}.
Figure \ref{fig:meanpt} shows $\langle p_T \rangle$
for pions
as a function of $T^{\mathrm{th}}$
at midrapidity $y=0$. 
In chemical equilibrium, $\langle p_T \rangle$
increases with decreasing $T^{\mathrm{th}}$.
On the contrary, $\langle p_T \rangle$
gradually decreases with decreasing $T^{\mathrm{th}}$
when early chemical freezeout is taken into account.
Even in the case that a
full hydrodynamic simulation without boost invariant
ansatz is performed and that
the contribution from resonance decays is included,
the temporal behavior of $\langle p_T \rangle$
as already discussed in Sec.~\ref{sec:ET}
is seen in Fig.~\ref{fig:meanpt}.
It should be emphasized here that
increase of $\langle p_T \rangle$
in chemical equilibrium
is a direct consequence of neglecting
chemical freezeout in hydrodynamic calculations
and, more definitely,
of the experimental results of particle ratios.
One has made full use 
of this unrealistic behavior
to reproduce $p_T$ spectra 
at the cost of hadron ratios
in the conventional
hydrodynamic calculations.

%%%%%%%%%%%%%%%%%%%%%%%%%%%%%%%%%%%%%%%%%%%%%%%%%%%%%%%%%%%%
% v2
%%%%%%%%%%%%%%%%%%%%%%%%%%%%%%%%%%%%%%%%%%%%%%%%%%%%%%%%%%%%
\begin{figure}[t]
\begin{center}
\includegraphics[width=3.3in]{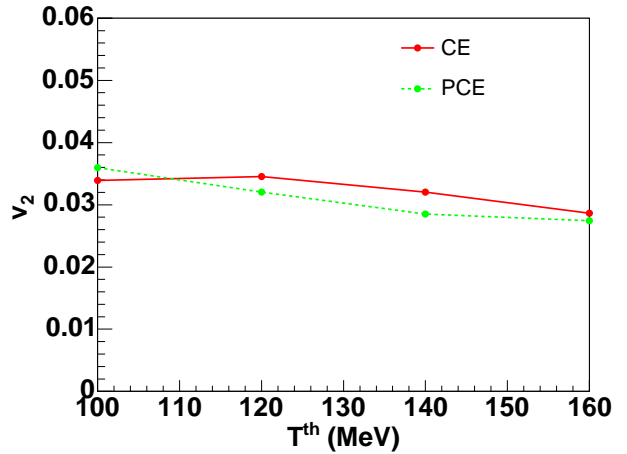}
\caption{
$v_2$ for pions as a function of $T^{\mathrm{th}}$
for the models CE (solid) and PCE (dashed).
}
\label{fig:v2}
\end{center}
\end{figure}
%%%%%%%%%%%%%%%%%%%%%%%%%%%%%%%%%%%%%%%%%%%%%%%%%%%%%%%%%%%%

In chemical equilibrium,
the slope of elliptic flow parameter $dv_2(p_T)/dp_T$
for pions
is insensitive to $T^{\mathrm{th}}$
and is consistent with the experimental data.
See also Figs.~8 and 11 in Ref.~\cite{Hirano:2002ds}.
This is apparently plausible since
the elliptic flow is 
a self-quenching phenomenon
and is sensitive to the early stage of the collisions
\cite{Ollitrault:1992bk}.
On the other hand, $dv_2(p_T)/dp_T$ for pions
in the model PCE
increases with decreasing $T^{\mathrm{th}}$
and is ended with overprediction of the experimental data
when $T^{\mathrm{th}}$ is chosen so as to reproduce the
proton $p_T$ spectrum and $v_2(p_T)$ in the low $p_T$ region.
This means that $v_2(p_T)$ for pions
varies in the late hadronic stage
($\tau\gton$ 10 fm/$c$).

We have to be careful in understanding
the difference between 
\textit{integrated} elliptic flow $v_2$
and \textit{differential} elliptic flow $v_2(p_T)$.
$v_2$ reflects the momentum anisotropy of bulk matter,
while $v_2(p_T)$ represents
how total $v_2$ distributes among various
particles with different $p_T$.
As shown in Fig.~\ref{fig:v2},
the integrated $v_2$ varies only weakly
with decrease of $T^{\mathrm{th}}$ 
(and hence proper time $\tau$)
in both cases.
This is consistent with the self-quenching picture 
of elliptic flow again.

\begin{figure}[t]
\begin{center}
\includegraphics[width=3.3in]{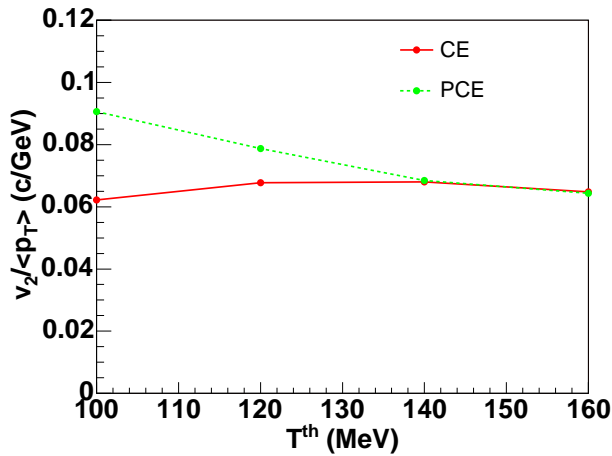}
\caption{
$v_2/\langle p_T \rangle$ for pions
 as a function of $T^{\mathrm{th}}$
for the models CE (solid) and PCE (dashed).
}
\label{fig:v2pttth}
\end{center}
\end{figure}

The slope of 
$v_2(p_T)$, on the other hand,  can be evaluated
approximately by 
$dv_2(p_T)/dp_T \approx v_2/\langle p_T \rangle $
since $v_2(p_T)$ for pions is almost a linear function of $p_T$
\cite{noteonv2slope}.
In chemical equilibrium (CE),
the moderate increase of $v_2$ ($\sim$13\% increase 
as $T^{\mathrm{th}}$ decreases from
160 MeV to 100 MeV)
is approximately canceled
by the increase of $\langle p_T \rangle$
($\sim$22\% from 160 MeV to 100 MeV).
 Eventually, the ratio
remains almost constant (or even decreases slightly)
as shown in Fig.~\ref{fig:v2pttth}.
The CE predictions for the differential elliptic flow
work because without chemical freezeout
the slope of $v_2(p_T)$ stalls accidentally and
reproduces the experimental data.

On the contrary, in PCE
the numerator $v_2$ increases by $\sim$20\%
and the denominator
$\langle p_T \rangle$ \textit{decreases}
by $\sim$10\%.
The resultant ratio $v_2/\langle p_T \rangle$ therefore 
increases with decreasing $T^{\mathrm{th}}$
as shown in Fig.~\ref{fig:v2pttth}.
This is the reason why the slope of $v_2(p_T)$ in PCE
increases even in the late stage after the spatial azimuthal asymmetry has 
reversed sign.

It is now easy to understand why 
$v_2(p_T)$ at the SPS energies is so close 
to the one at the RHIC energy (see, e.g. Fig.~17 in
Ref.~\cite{Adams:2005dq}).
This is due to the correlated change of
both $v_2$ and $\langle p_T \rangle$
from SPS to RHIC energies:
The increase of the average $v_2$  
is compensated for by the increase of $\langle p_T \rangle$.
The slopes of $v_2(p_T)$ therefore vary surprisingly weakly
with the beam energy.

%%%%%%%%%%%%%%%%%%%%%%%%%%%%%%%%%%%%%%%%%%%%%%%%%%%%%%%%%%%%
% v2/<pT>
%%%%%%%%%%%%%%%%%%%%%%%%%%%%%%%%%%%%%%%%%%%%%%%%%%%%%%%%%%%%
%%%%%%%%%%%%%%%%%%%%%%%%%%%%%%%%%%%%%%%%%%%%%%%%%%%%%%%%%%%%

Summarizing the discussion, differential elliptic flow 
\textit{is} sensitive to the late hadronic stage
in ideal hydrodynamic calculations with
early chemical freezeout since
the slope of $v_2(p_T)$ is determined by
the mean $p_T$, \textit{i.e.} radial flow.
The apparent consistency of $v_2(p_T)$
between RHIC data and
results based on conventional CE hydrodynamic simulations
is therefore fortuitous.

\section{Analytic Approach}
\label{sec:analytic}

In order to understand the effect of chemical freezeout
on the transverse momentum dependence of elliptic flow
analytically, we employ the  blast wave model
discussed in Ref.~\cite{Pratt:2004zq}.
In the framework of the blast wave model,
one can choose the radial flow parameter
and temperature independently to reproduce the 
slope of $p_T$ spectra.
However, these two parameters are correlated 
in actual hydrodynamic simulations.
After chemical freezeout, the mean $p_T$ decreases with
decreasing $T^{\mathrm{th}}$, i.e. $d\langle p_T \rangle /dT >0$,
as already discussed in
the previous sections. 
On the contrary, 
the mean $p_T$ increases with
decreasing $T^{\mathrm{th}}$, i.e. $d\langle p_T \rangle /dT <0$,
in chemical equilibrium.
So we can
constrain the average flow velocity as a function of temperature
through the condition $d\langle p_T \rangle /dT =0$.
We call the obtained radial flow the critical radial flow,
$v_r^{\mathrm{crit}}(T)$.
The critical radial flow ensures that the mean $p_T$
is a constant of motion.
Qualitatively, $v_r(T)<v_r^{\mathrm{crit}}(T)$ corresponds to
the chemically frozen system,
while $v_r(T)>v_r^{\mathrm{crit}}(T)$ corresponds to
the system under chemical equilibrium.
In the following in this section, we assume only pions 
which are dominant particles in a fluid element.

\subsection{Momentum Distribution}
The invariant momentum spectrum
is given by the Cooper-Frye formula \cite{Cooper:1974mv}
in the Boltzmann approximation:
\begin{eqnarray}
\label{eq:CF}
E\frac{dN}{d^3p} & = & \frac{g}{(2\pi)^3}\int p^\mu d\sigma_\mu
\exp\left(-\frac{p^\mu u_\mu-\mu}{T}\right).
\end{eqnarray}
Here $g$ is the degree of freedom,
$d\sigma^\mu$ is the element of freezeout hypersurface.
$p^\mu$ is the four momentum measured in the laboratory system.
$\mu$ and $T$ are, respectively, chemical potential and temperature
at freezeout.
Four fluid velocity ($u^\mu u_\mu = 1$) 
can be parametrized as
\begin{eqnarray}
u^\mu & = & \cosh\rho (\cosh\eta_f, \sinh\rho \cos\phi,
\sinh\rho \sin\phi, \sinh\eta_f). \nonumber \\
\end{eqnarray}
Here $\rho$ is the transverse rapidity and
$\eta_f$ is the longitudinal rapidity
which is to be identified with the space-time rapidity $\eta_s$
in the Bjorken boost invariant solution \cite{Bjorken:1982qr}.
One can also write $\cosh\rho=\sqrt{1+u_\perp^2}$ and $\sinh\rho=u_\perp$,
respectively.
Energy of a particle in the local rest frame becomes
\begin{eqnarray*}
p^\mu u_\mu = m_T \cosh(y-\eta_f)\sqrt{1+u_\perp^2}
-p_T u_\perp \cos(\phi_p -\phi).
\end{eqnarray*}
Here $y$ is the momentum rapidity and $\phi_p$ is the azimuthal
angle of the momentum.
According to Ref.~\cite{Pratt:2004zq},
we also make the same ansatz
$u_\perp(\phi) = u_\perp(1+\varepsilon\cos2\phi)$ 
for azimuthal dependence of radial flow and
take terms up to the first order in $\varepsilon$
\begin{eqnarray}
p^\mu u_\mu & = & m_T \cosh(y-\eta_f)\sqrt{1+u_\perp^2}
\left[1+\frac{u_\perp^2}{1+u_\perp^2}
\varepsilon\cos2\phi\right]\nonumber \\
&-& p_T u_\perp \cos(\phi_p -\phi) (1+\varepsilon \cos 2\phi).
\end{eqnarray}
Assuming the matter suddenly freezes out at $\tau_f$,
\begin{eqnarray}
p^\mu d\sigma_\mu = E dV
= m_T\cosh y \times rdrd\phi \tau_f d\eta_s .
\end{eqnarray}
Note that 
$u_\perp(\phi)$ is supposed to include all possible
azimuthal anisotropic effects
and that $\phi$ dependences of $T$ and $\mu$
are neglected as in the hydrodynamic approach.
Then Eq.~(\ref{eq:CF}) reduces to
\begin{eqnarray}
\frac{dN}{m_Tdm_T d\phi_p dy} & \propto & 
\int d\phi m_T \cosh y e^A,
\end{eqnarray}
\begin{eqnarray}
A & = & -\frac{m_T \cosh(y-\eta_f)\sqrt{1+u_\perp^2}}{T}\nonumber \\
& + & \frac{p_T u_\perp \cos(\phi_p-\phi)}{T}+\frac{\mu}{T}\nonumber \\
& - & \varepsilon\frac{m_T u_\perp^2 \cosh(y-\eta_f)\cos 2\phi}
{T\sqrt{1+u_\perp^2}} \nonumber \\
& + & \varepsilon
\frac{p_T u_\perp \cos 2\phi\cos(\phi_p-\phi)}{T}.
\end{eqnarray}
\subsection{Azimuthal Anisotropy}
By using the above momentum distribution,
one can calculate
$v_2(m_T)$ (or $v_2(p_T)$)
\begin{eqnarray}
\label{eq:v2}
v_2(m_T) = \frac{\int d\phi_p \cos 2\phi_p \frac{dN}{m_Tdm_Td\phi_p}}
{\int d\phi_p \frac{dN}{m_Tdm_Td\phi_p}}.
\end{eqnarray}
Thus we obtain the same equation as Eq.~(33) in Ref.~\cite{Pratt:2004zq}
\begin{eqnarray}
\label{eq:v2mt}
v_2(m_T) 
& = & \frac{\varepsilon}{J_0}
\left(-\frac{m_T u_\perp^2}{T\sqrt{1+u_\perp^2}}J_E
+ \frac{p_T u_\perp}{T}J_p\right),\\
\label{eq:j0}
J_0 & = & 2K_1 I_0, \\
J_E & = & \left(K_0 + \frac{K_1}{z_E}\right)I_2, \\
J_p & = & \frac{1}{2}\left(I_1 + I_3\right)K_1.
\end{eqnarray}
Here $K_i$ and $I_i$ are
modified Bessel functions.
It is always understood that the argument of $K_i$ ($I_i$)
is $z_E = m_T\sqrt{1+u_\perp^2}/T$
($z_p = p_Tu_\perp/T$).
Detailed calculations can be found in Appendix A.

One can also obtain an analytic expression for
the slope of $v_2(p_T)$
\begin{eqnarray}
\frac{dv_2}{dp_T}
& = & \frac{\varepsilon}{J_0^2}
\left\{-\frac{u_\perp^2}{T\sqrt{1+u_\perp^2}}
\left[\frac{p_T}{m_T}J_E \right.\right. \nonumber \\
 &- &\left.\left. m_T\left(\frac{\partial z_E}{\partial p_T}
\frac{\partial J_E}{\partial z_E}
+\frac{\partial z_p}{\partial p_T}
\frac{\partial J_E}{\partial z_p}
\right)\right]\right.\nonumber \\
 & +& \left.\frac{u_\perp}{T}J_p 
+ \frac{p_T u_\perp}{T}
\left(\frac{\partial z_E}{\partial p_T}
\frac{\partial J_E}{\partial z_E}
+\frac{\partial z_p}{\partial p_T}
\frac{\partial J_p}{\partial z_p}
\right)\right\}.\nonumber \\
\end{eqnarray}
Here,
\begin{eqnarray}
\frac{\partial z_E}{\partial p_T} & = &
\frac{\sqrt{1+u_\perp^2}}{T}\frac{p_T}{m_T},\\
\frac{\partial z_p}{\partial p_T} & = &
\frac{u_\perp}{T},\\
\frac{\partial J_E}{\partial z_E} & = &
-\left(K_1 + \frac{K_2}{z_E}\right)I_2,\\
\frac{\partial J_E}{\partial z_p} & = &
\frac{1}{2}\left(K_0 + \frac{K_1}{z_E}\right)(I_1+I_3),\\
\frac{\partial J_p}{\partial z_E} & = &
-\frac{1}{2}\left(K_0 + \frac{K_1}{z_E}\right)(I_1+I_3),\\
\frac{\partial J_p}{\partial z_p} & = &
\frac{1}{4}K_1(I_0 +2I_2+I_4).
\end{eqnarray}
Note that one can replace a higher order modified Bessel function
with lower order functions.

\subsection{Incorporation of Transverse Dynamics}

From discussion in Sec.~\ref{sec:ET},
we find
$d\langle m_T \rangle/dT < 0$
(or $d\langle m_T \rangle/d\beta > 0$)
for chemical equilibrium pions
and
$d\langle m_T \rangle/dT > 0$
(or $d\langle m_T \rangle/d\beta < 0$) for chemically frozen pions.
These features are quite generic for ideal Bjorken fluids of pions.
Obviously, the analytic approach is just a parametrization
at freezeout and contains almost no information about
the time evolution of the system.
For example, the analytic approach
does not tell us anything about how $u_\perp$ increases
with decrease of temperature.
In this subsection, we try to give a dynamical meaning to
the blast wave approach discussed in the previous subsection.

The transverse mass distribution in the analytic approach is
(see Eq.~(\ref{eq:j0})) \cite{Schnedermann:1992hp}
\begin{eqnarray}
\frac{dN}{m_T dm_T} & \propto & m_T K_1 I_0.
\end{eqnarray}
Thus one obtain the mean transverse mass
\begin{eqnarray}
\langle m_T \rangle = \frac{\int dm_T m_T^3 K_1
I_0}
{\int dm_T m_T^2 K_1I_0}
\end{eqnarray}
and its derivative
with respect to the inverse temperature
\begin{eqnarray}
\label{eq:meanmt_beta}
\frac{d\langle m_T \rangle}{d\beta} & = & \frac{1}
{\left(\int dm_T m_T^2 K_1 I_0 \right)^2}\nonumber \\
& &\times\left\{\left[\int dm_1 m_1^3
\left(\frac{dK_1}{d\beta}I_0
+ K_1\frac{dI_0}{d\beta}\right)\right]\right.\nonumber \\
& & \times \left(\int dm_2 m_2^2 K_1 I_0 \right)
-\left(\int dm_1 m_1^3 K_1 I_0\right)\nonumber \\
& & \times \left.\left[\int dm_2 m_2^2 \left(\frac{dK_1}{d\beta}I_0
+ K_1\frac{dI_0}{d\beta}\right)\right]
\right\},\\
\frac{dK_1}{d\beta} & = & \frac{\partial K_1}{\partial \beta}
+\frac{d\rho}{d\beta}\frac{dK_1}{d\rho} \nonumber \\
& = & m_T \frac{dK_1(z_E)}{dz_E}\left(\cosh\rho
+\frac{d\rho}{d\beta}\beta\sinh\rho\right) \nonumber \\
& = & -m_T\left(K_0+\frac{K_1}{z_E}\right)\nonumber \\
& & \times\left(\cosh\rho
+\frac{d\rho}{d\beta}\beta\sinh\rho\right),\\
\frac{dI_0}{d\beta} 
& = & p_T \frac{dI_0(z_p)}{dz_p}\left(\sinh\rho
+\frac{d\rho}{d\beta}\beta\cosh\rho\right) \nonumber \\
& = & p_T I_1\left(\sinh\rho
+\frac{d\rho}{d\beta}\beta\cosh\rho\right).
 \end{eqnarray}
 The numerator of Eq.~(\ref{eq:meanmt_beta}) reduces to
\begin{eqnarray}
&&\mathrm{[numerator \enskip of \enskip Eq.~(\ref{eq:meanmt_beta})]}\nonumber \\
 & = &\int dm_1 \int dm_2 m_1^2 m_2^2 (m_2-m_1)
(K_1I_0) \mid_{m_T=m_1}\nonumber\\
& \times &\left[\left(m_T\cosh\rho
+m_T\sinh\rho\frac{d\rho}{d\beta}\beta\right)
K_0 I_0 \right.\nonumber \\
&&+\left(\frac{1}{\beta}+\frac{d\rho}{d\beta} \tanh\rho\right)
K_1 I_0 \nonumber \\ 
&&-\left.\left.\left(p_T \sinh\rho+p_T\cosh\rho\frac{d\rho}{d\beta}\beta\right)
K_1 I_1\right]\right|_{m_T=m_2}. \nonumber \\
\end{eqnarray}
The second term in the 
square bracket vanishes due to the antisymmetry ($m_1 \leftrightarrow m_2$). Thus,
\begin{eqnarray}
&&\mathrm{[numerator \enskip of \enskip Eq.~(\ref{eq:meanmt_beta})]}\nonumber \\
 & = &\int dm_1 \int dm_2 m_1^2 m_2^2 (m_2-m_1)
K_1(z_{E,1})I_0(z_{p,1})\nonumber\\
& \times &\left[m_2\left(\cosh\rho
+\sinh\rho\frac{d\rho}{d\beta}\beta\right)
K_0(z_{E,2}) I_0(z_{p,2}) \right.\nonumber \\
&-&\left.\sqrt{m_2^2-m^2}\left( \sinh\rho
+\cosh\rho\frac{d\rho}{d\beta}\beta\right)
K_1(z_{E,2}) I_1(z_{p,2})\right],\nonumber \\ 
\end{eqnarray}
where $z_{E,i} = z_E \mid_{m_T = m_i}$
and $z_{p,i} = z_p \mid_{p_T = \sqrt{m_i^2-m^2}}$.
The sign of $d\langle m_T\rangle/d\beta$ is determined by this equation.
One can obtain the relation between $\rho$ and $\beta$
by solving an equation $d\langle m_T \rangle/d\beta =0$:
\begin{eqnarray}
\label{eq:crit_cond}
&&\beta\frac{d\rho}{d\beta}[F(\rho)\sinh\rho-G(\rho)\cosh\rho]\nonumber \\
& + &[F(\rho)\cosh\rho-G(\rho)\sinh\rho]=0,
\end{eqnarray}
where,
\begin{eqnarray}
F(\rho)& =& \int dm_1 dm_2 (m_2-m_1)m_1^2 m_2^3 \nonumber \\
& \times & K_1(z_{E,1})I_0(z_{p,1})K_0(z_{E,2})I_0(z_{p,2}),\\
G(\rho)& = & \int dm_1 dm_2 (m_2-m_1)m_1^2 m_2^2 \sqrt{m_2^2-m^2}\nonumber \\
& \times & K_1(z_{E,1})I_0(z_{p,1})K_1(z_{E,2})I_1(z_{p,2}).
\end{eqnarray}
%%%%%%%%%%%%%%%%%%%%%%%%%%%%%%%%%%%%%%%%%%%%%%%%%%%%%%%%%%%%
% Critical line
%%%%%%%%%%%%%%%%%%%%%%%%%%%%%%%%%%%%%%%%%%%%%%%%%%%%%%%%%%%%
\begin{figure}[t]
\begin{center}
\includegraphics[width=3.3in]{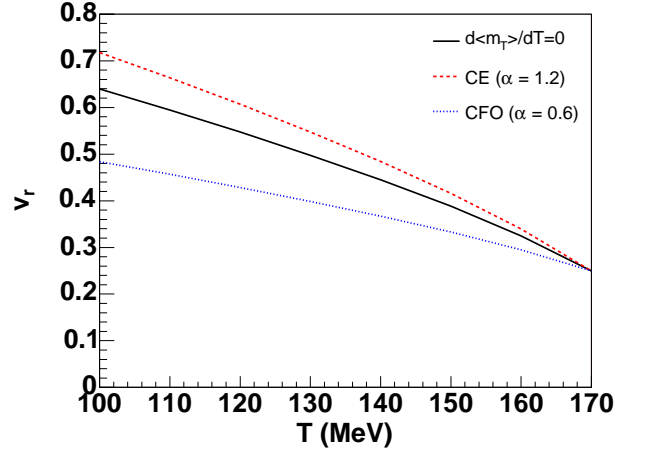}
\caption{
Temperature dependences of radial flow in the hadron phase.
Solid line shows a solution of 
the critical flow (see text for details)
with a given initial condition $(T,v_{r})=(170, 0.25)$.
Examples for the model CE ($\alpha=1.2$, dashed line)
and CFO ($\alpha=0.6$, dotted line)
are shown.
}
\label{fig:vtt}
\end{center}
\end{figure}
%%%%%%%%%%%%%%%%%%%%%%%%%%%%%%%%%%%%%%%%%%%%%%%%%%%%%%%%%%%%
The temperature dependence of transverse rapidity $\rho=\rho(\beta)$
is obtained for a given ``initial" condition $(\beta_0,\rho_0)$.
This particular radial flow ensures the mean transverse mass
becomes a constant and is an upper limit
of average
 radial flow in chemical freezeout EOS for massive pions.
We call this solution  the critical radial flow
$v_{r}^{\mathrm{crit}} = \tanh \rho(\beta)$.
One can parametrize the temperature dependence of
radial flow by introducing a parameter $\alpha$ 
within the analytic approach
which embodies the transverse dynamics
of the chemically frozen/equilibrated pion fluid:
\begin{eqnarray}
v_{r} (T)  =  v_{r}(T^{\mathrm{ch}})
 + \alpha [v_{r}^{\mathrm{crit}}(T)-v_{r}(T^{\mathrm{ch}})],
\end{eqnarray}
where $v_{r}(T^{\mathrm{ch}}) = \tanh\rho_0(\beta_0)$
is an initial condition
for Eq.~(\ref{eq:crit_cond}).
Although the exact value of $\alpha$ needs much more involved
dynamical calculation, radial flow qualitatively
corresponds to the chemically frozen system for $0<\alpha<1$
and to the chemical equilibrium system for $\alpha>1$.
It should be mentioned here that the temperature 
dependence of average
radial flow can be described to some extent without
solving hydrodynamic equations.
Note that $\alpha$ should be taken as being a moderate value
so that the total energy of the system (per unit rapidity) 
does not increase due to generation of radial flow.

Figure \ref{fig:vtt} shows temperature dependences of
radial flow. 
The solid line shows the critical radial flow
$v_r^{\mathrm{crit}}$
which results from $d\langle m_T \rangle/dT = 0$.
This is obtained by solving Eq.~(\ref{eq:crit_cond}) with
an initial condition $(T,v_{r})=(170, 0.25)$
which is consistent with
a value at RHIC energies \cite{Hirano:2002ds}.
The critical flow distinguishes the system of
chemical equilibrium from that of chemical freezeout:
$v_{r}(T)$ for CE
(CFO) should be located above (below) this line
since $d\langle m_T \rangle/dT < 0$
for CE
($d\langle m_T \rangle/dT > 0$ for CFO).
Dashed line ($\alpha=1.2$) shows an example of
radial flow in the analytic model CE,
while dotted line ($\alpha=0.6$) shows the one
in the analytic model CFO.
These results look very similar to the results from
real hydrodynamic simulations as shown in Fig.~5 
in Ref.~\cite{Hirano:2002ds}.

%%%%%%%%%%%%%%%%%%%%%%%%%%%%%%%%%%%%%%%%%%%%%%%%%%%%%%%%%%%%
% v2(pT) from analytic model
%%%%%%%%%%%%%%%%%%%%%%%%%%%%%%%%%%%%%%%%%%%%%%%%%%%%%%%%%%%%
\begin{figure}[t]
\begin{center}
\includegraphics[width=3.3in]{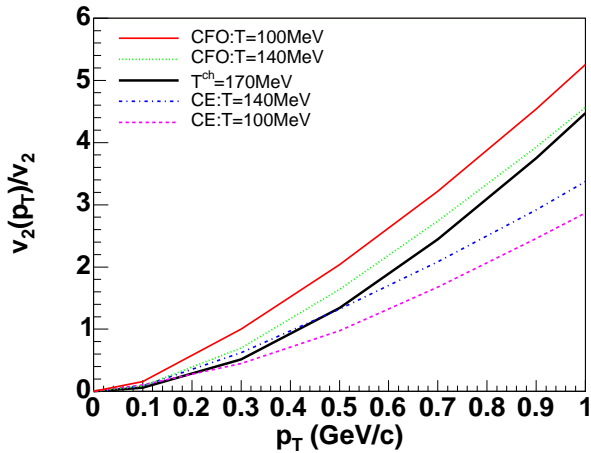}
\caption{
Elliptic flow parameter as a function of transverse momentum
within the analytic approach for pions.
Thick solid line shows $v_2(p_T)/v_2$ at chemical freezeout.
Thin solid (dotted) line shows the result from the 
analytic model CFO
($\alpha = 0.6$) at $T^{\mathrm{th}}=$100 (140) MeV.
Dashed  (dashed-dotted) line shows the result from the 
analytic model CE
($\alpha = 1.2$)
at $T^{\mathrm{th}}=$100 (140) MeV.
}
\label{fig:v2pt}
\end{center}
\end{figure}
%%%%%%%%%%%%%%%%%%%%%%%%%%%%%%%%%%%%%%%%%%%%%%%%%%%%%%%%%%%%

By using these profiles for radial flow,
we calculate $v_2(p_T)$ for pions
below the chemical freezeout
temperature.
Hydrodynamic analysis tells us that the \textit{integrated} $v_2$
is saturated within first 3--4 fm/$c$ just after the collision
and 
insensitive to the late hadronic stage \cite{Kolb:2003dz}.
However, within our analytic approach, there is no 
dynamical mechanism
which saturates the \textit{integrated} $v_2$
in the late hadron stage.
Therefore $v_2(p_T)/v_2$ is the quantity
to be compared with the results
from full hydrodynamic simulations.

In Fig.~\ref{fig:v2pt},
$v_2(p_T)/v_2$ for the analytic models CE and CFO
are represented.
The thick solid line shows the result
at $T=T^{\mathrm{ch}}=170$ MeV.
The overall slope up to 1 GeV/$c$ is gradually increasing with
decreasing the temperature in the analytic model CFO,
while the slope is decreasing in the analytic model CE.
These results clearly show that
the slope of $v_2(p_T)$ can vary
in the late hadronic stage and that
the temporal behavior of average transverse momentum
$\langle p_T \rangle$ and 
radial flow $v_r$
is the key to understand the shape
of $v_2(p_T)$.

\section{Conclusion and Outlook}
\label{sec:summary}
In this paper we showed that the  differential elliptic flow
observable $v_2(p_T)$, which is a critical component
for the interpretation of RHIC data
in terms of perfect fluidity of the sQGP core,
 is sensitive to the degree of hadro-chemical equilibrium
in late time evolution of the hadronic corona.
If local equilibrium hydrodynamics 
is applied to the hadronic corona {below}
$T_c$, an inevitable logical {impasse} arises when confronting
all the data on (1) hadron abundances (2) radial flow and (3) differential
elliptic flow. In CE hydrodynamics (2) and (3) can be reproduced
at the expense of (1). In PCE (1) is enforced at the expense
of (2) and (3) as summarized in Table I. We presented
a simple analytic
blast wave model to explain these nonintuitive consequences
of hadro-chemical (non)equilibrium in (P)CE implementations of
hydrodynamics.

In CE both 
the average transverse momentum per hadron $\langle p_T\rangle$ 
and average $v_2$ increase with proper
time in the hadronic phase 
in a way that \textit{accidentally} preserves
the slope of
 differential elliptic flow
$dv_2(p_T)/dp_T\approx v_2/\langle p_T\rangle$ in agreement with the data.
In PCE,  $\langle p_T\rangle$ \textit{decreases} due to the basic Bjorken
longitudinal cooling. The main point is that
in PCE the hadronic yields are fixed at $T^{\mathrm{ch}}$ 
and the compensating CE
``local reheating'' mechanism (the conversion of heavy resonance mass
back into internal energy
which ``mimics" a sort of dissipative effect)
is absent.
This is why PCE fails to describe
the proton radial flow data. In addition,
the slight increase of the average
$v_2$, as in CE, with proper
time cannot be compensated for in PCE. Therefore,
 {the slope of \textit{differential} elliptic flow
$dv_2(p_T)/dp_T\approx v_2/\langle p_T\rangle$} continues to grow in PCE 
during the hadronic phase,
which leads to disagreement with RHIC data.

The subtle interplay {among} (1) longitudinal expansion work,
(2) maintenance of hadronic abundance yields, (3) the long time development
of radial flow, and (4) the differential azimuthal asymmetric elliptic flow
provides a formidable dynamical constraint on the dynamics
of the hadronic corona. Only by abandoning ideal hydrodynamics
in the hadronic corona, have nonequilibrium hadron cascade (HC) models 
been
able to deal with the interplay of the above hadron dynamics
in a way consistent with present RHIC data.
As discussed in Sec.~\ref{sec:etaovers}, this approach is natural 
since the viscosity to entropy
ratio in a hadronic resonance gas below $T_c$ is too large to support 
even local thermal equilibrium.
By relaxing both thermal and hadro-chemical equilibrium assumptions, 
the hybrid QGP {hydrodynamics} plus hadron cascade
model in \cite{Teaney:2000cw} has been able to account for all three
major low $p_T$ observables as summarized in Table I. 
The effect of viscosity in the hadron phase
\cite{Teaney:2003pb}
substitutes for the ``local reheating'' in the CE model
and compensates the small growth
of the average $v_2$ in PCE to preserve
the slope of $v_2(p_T)$ for pions.
The slope of $v_2(p_T)$ is thus found to stall
at the SPS energy
from the hybrid model analysis \cite{Teaney:2002aj}.
In the classical transport approach,
both $\langle p_T \rangle$ \cite{Gyulassy:1997ib}
and $v_2$ \cite{Zhang:1999rs}
do not vary significantly
when the mean free path among the particles becomes
comparable with the typical gradient length scales.
Moreover, the shear viscous effect changes the momentum
distribution function \cite{Dumitru:2002sq}
 and reduces the slope of $v_2(p_T)$
slightly \cite{Teaney:2003pb}.
These are the reasons why
the slope of $v_2(p_T)$
does not changed significantly
in the hadronic transport stage.

So how robust is the statement that
hydrodynamic description at RHIC 
works remarkably well?
We emphasized that the behavior of $v_2$
differs from that of $v_2(p_T)$: The integrated elliptic
flow does not develop so much in the late hadronic stage
in which either the inviscid, chemical (non-)equilibrium fluid
or the dissipative gas is assumed,
whereas the differential elliptic flow depends largely on
these assumptions.
The large magnitude of \textit{integrated} $v_2$
observed at RHIC is reproduced
only when a small $\eta/s$ is assumed \cite{Molnar:2001ux}.
Therefore the large $v_2$ developed in the early stage
is obtained from
the evolution of the sQGP core, which as discussed in 
Sec.~\ref{sec:etaovers} must have near  minimal viscosity
$\eta_{\mathrm{SYM}}\approx T^3$.
Even though the minimal sQGP viscosity is larger than the 
viscosity of the HRG corona, the core exhibits near perfect fluid
behavior due to the deconfinement of almost all
the QCD degrees of freedom. The near perfect fluidity of the sQGP core
is therefore a signal of deconfinement. 
On the other hand, the breaking of local and hadro-chemical
equilibrium in the hadronic corona is critical
for this interpretation of RHIC data.
If inviscid ideal hydrodynamics were valid in both
sQGP and HRG phases, the crucial $v_2(p_T)$ would be
sensitive to the hadronic thermal freezeout dynamics
and not only to the equation of state of sQGP matter.

Perhaps most surprising in connection with the hydrodynamics
robustness question
is the important role played
by hadro-chemical freezeout at $T^{\mathrm{ch}}\approx T_c$
that is implied by the extensive systematics of
observed hadron abundances \cite{Braun-Munzinger:2003zd}.
Without this constraint the different hadro-chemical
results with  CE and PCE 
would preclude a conclusion about the perfect fluidity of the sQGP
core as well as 
the highly dissipative, imperfect fluidity
of its hadronic resonance corona.
 
Despite the success of the hybrid HC approach, there exist 
open technical questions
that must be still investigated.
One important issue is
the violation of energy-momentum conservations
at the boundary between the QGP and hadron phases \cite{Bugaev:2002ch}.
The Cooper-Frye prescription \cite{Cooper:1974mv}
is employed to obtain the particle distribution
just after hadronization 
which is to be used as an initial condition
in the sequential cascade calculation.
{The} {violation} {is expected to be small
when radial flow is large. Nevertheless} 
there always exists in the space-like hypersurface $d\sigma^\mu$
 in-coming particles which contributes
to the number of particles negatively.
This negative contribution is omitted in the actual calculations,
which causes the violation of energy-momentum conservation.
Proper treatment of the boundary condition may lead to change
the dynamics in the QGP phase \cite{Bugaev:2002ch}.
In this connection, the approximate continuity of the viscosity from the sQGP 
to the HRG phase discussed in Sec.~\ref{sec:etaovers} minimizes this
interface problem since there is no discontinuity of
the stress tensor including the viscous correction at $T_c$.

Another important future problem
is the rapidity dependence of elliptic flow.
The (3+1)-dimensional ideal hydrodynamic
calculations \cite{Hirano:2001eu,Hirano:2002ds}
have not been
able to reproduce the observed pseudorapidity
dependence of $v_2$
in forward rapidity region at RHIC
\cite{Back:2002gz}.
The forward rapidity region at RHIC
is similar to the midrapidity region at SPS
in the sense that local particle density $(1/S)dN_{\mathrm{ch}}/dy$
is similar.
Heinz and Kolb \cite{Heinz:2004et}
proposed 
a ``thermalization coefficient"
to describe enhanced nonequilibrium effects
in the low particle density region
defined from 
the experimental data $v_2/\varepsilon$
as a function of $(1/S)dN_{\mathrm{ch}}/dy$ 
\cite{Alt:2003ab}.
To address correctly the rapidity and beam energy dependence
{taking into account} the highly viscous nature of the hadronic corona,
a new hybrid model must be developed
in which the (3+1)-D hydrodynamic model for the sQGP core
is combined with the HRG transport approach
for the dissipative hadronic corona.
Hirano and  Nara
\cite{Hirano:2004rs,Hirano:2003pw,Hirano:2002sc}
have already developed
a dynamical model
to describe three different aspects
of relativistic heavy ion collisions in one consistent framework:
Color glass condensate initial conditions for high energy 
colliding nuclei, hydrodynamic evolution in 3D space
for long wave length components of produced matter,
and quenching jets for short wave length components.
A further unified framework
by combining a hadronic cascade model
with the above one
\cite{HiranoNara}
will be necessary to understand more quantitatively
the dynamics and
the properties of QCD matter
produced in relativistic heavy ion collisions
at all SPS, RHIC and LHC energies.

We emphasize that continued work toward such a unified
dynamical framework will be  essential to further test
our physical interpretation of RHIC data -
as outlined in the introduction and analyzed in sections II-VI -
that ``perfect fluidity'' of the higher viscosity sQGP core
and ``imperfect fluidity'' of the lower viscosity HRG corona
taken together with hadro-chemical equilibrium near $T_c\sim 160-170$ MeV
already provide a compelling set of signatures for QCD deconfinement at RHIC. 

\begin{acknowledgements}
 
We thank
C.~Ogilvie for discussions
on compiled hydrodynamic results
on transverse momentum dependences
of spectra and elliptic flow in the PHENIX white paper.
Valuable discussions with K.~Bugaev,
P.~Huovinen, D.~Molnar, K.~Schalm and D.~Teaney
are also acknowledged.
We also thank P.~Huovinen for cereful reading of the manuscript.
This work was supported in part by the United States
Department of Energy
under Grant No.~DE-FG02-93ER40764.
\end{acknowledgements}

\vspace{12pt}

\appendix

\section{Derivation of Eq.~(\ref{eq:v2mt})}
Let us recall the formulae for modified Bessel functions
\begin{eqnarray}
2K_1(z) & = & \int_{-\infty}^{\infty}dx \cosh x \exp(-z\cosh x),\\ 
2\pi I_2(z) & = & \int_{-\pi}^\pi d\phi \cos 2\phi \exp(z\cos\phi). 
\end{eqnarray}
The numerator of Eq.~(\ref{eq:v2}) becomes
\begin{eqnarray}
& &\int d\phi_p \cos 2\phi_p
\int d\phi \int dy \cosh y \exp(A) \nonumber \\
& = & \int d\phi 2K_1(B)  \times 2\pi \cos 2\phi I_2(C),
\end{eqnarray}
where,
\begin{eqnarray}
&& B = \frac{m_T\sqrt{1+u_\perp^2}}{T}
+\varepsilon \frac{m_T u_\perp^2 \cos 2\phi}{T\sqrt{1+u_\perp^2}}, \nonumber \\
&& C = \frac{p_Tu_\perp}{T}
+\varepsilon \frac{p_T u_\perp \cos 2\phi}{T}. \nonumber
\end{eqnarray}

We here assume $\varepsilon$ is small, expand the modified
Bessel function, and take the first order term
with respect to $\varepsilon$,
\begin{eqnarray}
&& 4\pi \int d\phi \left(K_1
+\varepsilon K_1' \frac{m_T u_\perp^2 \cos 2\phi}
{T\sqrt{1+u_\perp^2}}\right)\nonumber \\
& \times & \left(I_2 + \varepsilon I_2' 
\frac{p_T u_\perp \cos 2\phi}{T} \right) \nonumber \\
& \approx & 4\pi^2 \varepsilon \left(K_1'I_2 
\frac{m_T u_\perp^2}{T\sqrt{1+u_\perp^2}} 
+K_1 I_2' \frac{p_T u_\perp}{T}\right).
\end{eqnarray}
Let us also recall some useful formulae
for the derivatives of modified Bessel functions,
\begin{eqnarray}
K_n'(z) & = & -K_{n-1}(z) -\frac{n}{z}K_n(z),\\
I_n'(z) & = & \frac{1}{2}\left[I_{n-1}(z) +I_{n+1}(z) \right].
\end{eqnarray}
Then the numerator of Eq.~(\ref{eq:v2}) is proportional to
\begin{eqnarray}
4\pi^2 \varepsilon
& &\left[ -  \left(K_0 + \frac{K_1}{z}\right)I_2 
\frac{m_T u_\perp^2}{T\sqrt{1+u_\perp^2}}\right.\nonumber \\
& & \left. +  \frac{1}{2}\left(I_1 + I_3\right)K_1 \frac{p_T u_\perp}{T}\right].
\end{eqnarray}
Here the arguments of $K$ and $I$ are, respectively, 
$z_E = m_T\sqrt{1+u_\perp^2}/T$
and $z_p = p_Tu_\perp/T$.
An analogous calculation can be done
for the denominator of Eq.~(\ref{eq:v2}).
The result is proportional to $4\pi^2 \times 2K_1 I_0$.


\begin{thebibliography}{99}
\bibitem{noteonv2slope}
If $v_2(p_T)$ is a linear function of $p_T$, i.e $v_2(p_T)=c p_T$,
the equality $c=dv_2(p_T)/dp_T = v_2/\langle p_T \rangle$
holds exactly.
\end{thebibliography}
\end{document}